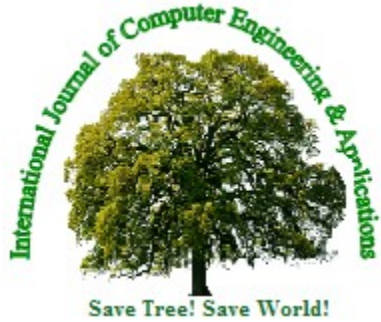

# A NOVEL ARCHITECTURE FOR RELEVANT BLOG PAGE IDENTIFICATION

**Deepti Kapri[1], Rosy Madaan[2], A. K Sharma[3], Ashutosh Dixit[4]**

[1]*M.Tech Scholar, Echelon Institute of Technology, Faridabad, Haryana INDIA*
[2] *Department of Computer Science & Engineering, Echelon Institute of Technology, Faridabad, India*
[3,4] *YMCA University of Science & Technology Faridabad, Haryana INDIA*

**ABSTRACT:**

**Blogs are undoubtedly the richest source of information available in cyberspace. Blogs can be of various natures i.e. personal blogs which contain posts on mixed issues or blogs can be domain specific which contains posts on particular topics, this is the reason, they offer wide variety of relevant information which is often focused. A general search engine gives back a huge collection of web pages which may or may not give correct answers, as web is the repository of information of all kinds and a user has to go through various documents before he gets what he was originally looking for, which is a very time consuming process. So, the search can be made more focused and accurate if it is limited to blogosphere instead of web pages. The reason being that the blogs are more focused in terms of information. So, User will only get related blogs in response to his query. These results will be then ranked according to our proposed method and are finally presented in front of user in descending order**

## [I] INTRODUCTION

World Wide Web is the repository that contains vast amount of information. It contains solution to almost every kind of user's query. However locating the relevant information to the user's query is a tedious task. On WWW, every user uses *search engine* for searching information about topics of interest. User enters his question and in response, gets a list of result pages which may serve the purpose.

A user enters its query on the interface. The query is at first preprocessed and is then passed to the searcher module that carries out the search in the index. The index is a mapping between the terms and the document(s) in which they appear. These are then ranked. The response is then passed back to search engine interface which in turn is then forwarded back to the user. It uses special software called *crawler*[10] which traverses the entire web and using the list of URLs downloads the web pages. The downloaded pages are then stored in a repository and are further scanned to extract the existing URLs in them. These URLs are then added to the list of seed URLs for further crawling.

Blogs originally came from the word "weblog" or a "web log" and the collection of blog pages on various topics is known as blogosphere. It is basically an informational site that consists of discrete entries called as *posts*. A web blog is a collection of blog posts that are arranged in reverse chronological order with the latest post appearing on the top and so on. Blogs can be of any kind like a political blog contains posts on political affairs, a sports blog containing sports information, a fashion blog contains posts on fashion respectively.. Blogs are more likely to be focused as they contain information on the topic being discussed. Any reader can interact with the blogger or can easily communicate with other fellow readers by providing its feedback or by posting comments, liking, voting or sharing the blog post

So, Blogs provide a very useful source of information that can be used to provide information to the user for their query. Blogs may contain some non-relevant information, so it becomes crucial to filter out such content and to provide only the relevant content to the user as per his needs. Some blogs may be more important than others. So, there is a need to identify the one that is most important. This is a big issue that arises.

Summarization is a technique that is used to extract those sentences from a web page that best represents its content in few lines. Formally, *Summary* [3] can be defined as a text that is produced from one or more texts, that contain a significant portion of the information in the original text(s), and that is no longer than half of the original text(s). The advantage of creating summary is that, user does not have to go through the entire content to find what the page is about. *Summarization* is the process of distilling the most important information from a source to build such a version that represents the whole text. So after the search within the blog summaries for the user's query, user is presented with the relevant content.

## [II] RELATED WORK

Justus Bross et. al [5] discussed various web page ranking algorithms. The paper compared the strengths and shortcomings of various algorithms. It deduced that they are only marginally applicable in case of blogs. Also, a novel ranking metric has been proposed called ''BLOGINTELLIGENCE-Impact- score'' or ''BI-Impact'' for short.

Apostolos Kritikopoulos et. al [6] discussed the limitation of traditional websites in giving proper results. An algorithm for calculating blog ranking has been proposed. The algorithm was basically an extension of page rank algorithm which analyses and extends the link graph, in an attempt to exploit some of the weblog features

M.A Tayebi et. al[9] discussed the traditional web page ranking algorithms and stated that they are insufficient for ranking blogs. In the paper a new ranking algorithm for ranking blogs using the user's behavioral features has been proposed.

Y. Jhu et al[11] proposed a novel ranking algorithm which uses both the link analysis and the content analysis of the blog. The algorithm considers more implicit features of blog, such as common topics, to improve the satisfaction of the users. Observing the previous work done on the above topic, we can deduct that Blogs undoubtedly are more preferred over general web pages for finding the accurate result. There were some limitations in these ranking algorithms:

1. Algorithms proposed by Apostolos and J Shen were based only on the link structure.
2. Other proposed algorithms were based on few behavioral features.
3. Even after limiting our search in blogosphere only, sometimes the result we got are not related with the query.

So, to make the search results more accurate and for identifying relevant blog pages, We have proposed an architecture given in the following section, thus providing the user better results to the user.

## [III] PROPOSED WORK

The paper proposes and discusses the architecture for searching in the index of blog summaries and provides the result in the form of list of pages. Also, the major work is done for identifying relevant blog pages. For this purpose a novel algorithm for ranking is proposed.

**Fig: 1 Proposed Architecture**

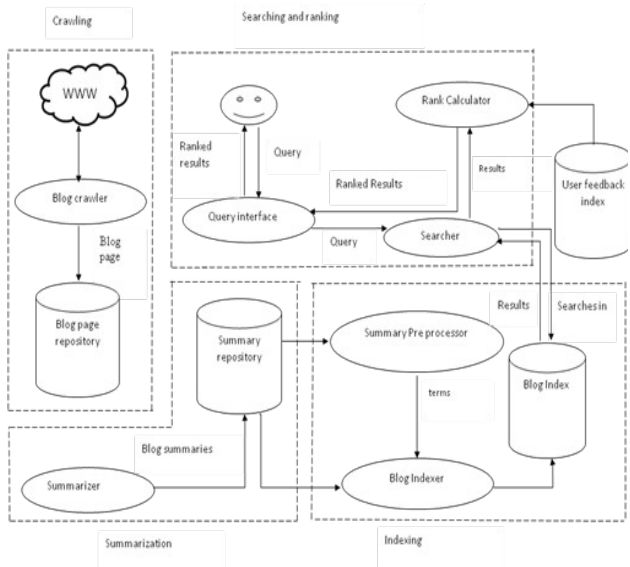

The architecture of the proposed work is given in fig.1. It mainly consists of following functional modules:

1) Blog Crawler
2) Summarizer
3) Summary pre processor
4) Blog Indexer
5) Searcher
6) Rank Calculator

### 3.1 Blog Crawler

Blog crawler is similar to a general crawler except that it restricts its crawl boundary to the blog space, thus downloading only the blog pages and ignoring rest of the web. The blog crawler differentiates the blog pages from the general web pages on the basis of the features [8] discussed below.

- A Blog word is found in the URLs of the blog pages.
- There is RSS tag.
- The most of the hyperlinks point to the blog itself.
- The blog posts are ordered by date, the recent one appears at the top.
- There are many other features of a blog page like the page includes author's information, comments, likes/dislikes, log archive and many others.

The authors in [8] proposed a novel architecture for the Blog crawler. The blog crawler crawls blog pages and stores them in a repository called as Blog page repository.

### 3.2 Summarizer

This module takes the blog pages stored in the Blog page repository and generates summary which is then stored in the summary repository. Summarizer [7] is a tool that is used to create a short description of any web page in two-three paragraphs. Summaries contain only the relevant data, giving a glimpse of underlying content to the user. This gives an idea to the user, what the page is all about and the user do not have to go through the entire web page. The summarization technique discussed in [7] creates blog heading and the content as well.

Summary is created on the basis of 2 factors PF and TF. PF i.e. *Presence factor* indicates the presence of blog heading terms, in the blog content. The sentences containing none of the blog heading terms will not be included in the summary. *Term frequency* indicates the frequency of the terms in blog heading in each sentence of the blog page. More the number of times a heading term appears in a sentence, more important is the sentence for the summary. Both factors are considered to score each sentence of the blog page. Then these sentences are ranked in descending order of their score, thus forming a summary.

### 3.3 Summary Preprocessor

This module takes the summaries from the summary repository and processes them. On processing, the pre-processor performs 3 main tasks. These are tokenization, stop word removal and stemming.

### 3.4 Blog Indexer

Blog Indexer takes the summaries from the summary repository and the terms generated after pre-processing and create a *blog index.* A blog index is basically an inverted index that is a mapping between terms and the blog summaries containing terms. The index contain *term frequency* i.e. number of times it is present, its *location* i.e. in heading section or body section, *summary ID* which indicates the id of the summary in which it is present. Table 1 is a sample index.

**[Table 1]**

| Terms | Info |
|---|---|
| Education | {5,<B>,B1} , {3,<H>,B2} , {1,<B>,B10} |
| Element | {1,<H>,B7} |
| Faridabad | {4,<B>,B2} , {2,<B>,B3} |
| Indian | {1,<B>,B1} , {3,<H>,B2} |
| Jammu | {3,<B>,B4} , {3,<H>,B5} |
| Politics | {7,<B>,B1} |
| Sachin | {2,<H>,B6} , {3,<H>,B8} |

Table: 1. Blog index

Ex For a term *Element*, information stored in blog index is {1,<H>, B7}. This information indicates that the term *Element* is present in summary id 7, in heading section and the term frequency is 1.

### 3.5 Searcher

The user enters his queries on the interface which is then passed to the searcher module. The searcher performs searching in the blog Index and returns the result to the ranking module for the purpose of ranking. The results returned by the module is the set of blog papers containing one or more query terms, frequency of each query term and the section (whether body or head) in which the term appears. Table 3 shows the results when searching for the query "Politics in Jammu".

### 3.6 Rank calculator

The searcher module passes the intermediate results to the rank calculator. This module will then rank each blog page by their blog scores. The results are then arranged accordingly in the descending order of their blog score. It has been found that there are two factors on the basis of which the blog page can be ranked.

2 main factors are i.e. *User Query factor* and *user feedback factor.*

*A) User Query* factor depends on the query terms entered by the user. Each term of the user's query plays a vital role in deciding the rank of any particular blog post. Presence of query term in a particular blog post indicates that, post may contain information answers relevant to user's query and hence might provide satisfactory response to user's query. Keeping this in mind, each blog post that contains the query terms will be included in the intermediate results. So whenever a user will enter his query, the searcher will search in the blog index shown in table 1, retrieves the summary id containing term(s) and also the information associated. The associated information tells where the term is present in the summary and how many times.

Presence of term in the heading (with <H> tag) is assigned more weight as compared to the term present in the body (with <B> tag)

Ex: If the user's query is, *"Computer Network"*, then after initial processing of the query, the result after searching in the inverted index will be {5,<B>,B1} and {1,<H>,B4}

For Blog B1: B1 contains only 1 term i.e. "*Computer*" in the body section (weight assigned=1), The term appears 5 times. So score for B1=5+1=6. Calculating average i.e. 6/2=3. Multiplying by its weight i.e. 0.6= 3x0.6=1.8.

For Blog B4: B4 contains only 1 term i.e. "*Network*" in the heading section (weight assigned=2), the term appears 1 times. So score for B1=1+2=3. Calculating average i.e. 3/2=1.5. Multiplying by its weight i.e. 0.6= 1.5x0.6=0.9

b)The second factor i.e. *User feedback factor* also contributes in the ranking of a blog page. In blogosphere, a user can interact with fellow readers or can provide his reviews in several ways. He can comment, vote or share the post on other platforms. So, there are 8 important parameters using which the user can provide his valuable feedback, which in turn increase the popularity of that post. These are as follows:

   i) Number of subscribers

   ii) Number of valid comment

   iii) Number of votes

   iv) Number of likes

   v) Rating

   vi) Sharing

   vii) Tagging

   viii) Presence of blogger information

*Number of subscribers* is the total number of people that have subscribed any web blog to their feeds. Each time any blogger publish new content on the blog, his followers receive an update, either on their Read Blogs page or via email

*Number of valid comments*: Readers can express their views or opinions by commenting on the blog posts. But it is often seen that, these comments are not related to the blog posts at all or are vague. So we will consider only valid comments here, which are related to the blog heading.

*Voting, Rating, Sharing* are some other ways in which user can review a blog. User can like the post, rate it on a scale of 5/10 as per his wish or can share it on various social networking platforms too like facebook, twitter, google+ etc.

*Blog tags* allow the author to place the blog into a particular category. Tags provide a useful way to group related posts together and to quickly tell readers what a post is about. Tags also make it easier for people to find your content.

Bloggers generally provide their information too in their home page.

Rank calculator module considers all these parameters for calculating the blog scores. We have considered eight parameters that are important in any blog page i.e. The first 6 factors are given a threshold of 0.3 and last 2 factors are given a threshold of 0.1.

Let's say weight assigned to F1 to F6 is 0.3
Let's say weight assigned to F7 to F8 is 0.1

**[Table 2]**

|         | F1    | F2              | F3     | F4     | F5      | F6       | F7    | F8           |
|---------|-------|-----------------|--------|--------|---------|----------|-------|--------------|
| Blog id | #subs | #Valid comments | #votes | #likes | #rating | #sharing | #tags | Blogger info |
| B1      | 2     | 0               | 4      | 0      | 0       | 0        | 4     | 1            |
| B2      | 10    | 3               | 4      | 0      | 7       | 56       | 3     | 0            |
| B3      | 19    | 4               | 10     | 30     | 3       | 3        | 0     | 1            |
| B4      | 14    | 10              | 20     | 3      | 3       | 2        | 5     | 1            |

| B5 | 12 | 3 | 0 | 0 | 0 | 0 | 4 | 0 |

Table: 2. User feedback Index

The Algorithm proposed for the ranking module is given below.

*Algorithm of Rank calculator Module*

*Input:* Blog Summaries from Searcher module say B1, B2; User feedback Index.

*Output:* Score of each blog summary i.e. SB1, SB2

*Algorithm:*

{

1) Calculate score of each Blog summary using "User Query factor" as follows:
   a) For each query term, add its frequency to its location in the summary.
   b) Calculate average score & multiply by the weight assigned to the factor.
   c) Return B1s, B2s and so on.
2) Calculate score of each Blog summary using "User Feedback factor" as follows:
   a) Retrieves factor F1 to F8.
      F1= # of subscribers
      F2=# of valid comments
      F3= # of votes
      F4= # of likes
      F5= # of rates
      F6= # of shares
      F7= # of tags
      F8= Blogger information
   b) Add factors from F1 to F6 calculate average and multiply by the weight assigned to the factor.
   c) Add factors from F7 to F8 calculate average and multiply by the weight assigned to the factor.
   d) Return blog feedback score i.e. B1fs, B2fs and so on.
3) Overall Blog Score i.e. SB1, SB2 is calculated by adding the scores obtained in 1 and 2. i.e. SB1=B1s+B1fs and returned back to the user}

Let the user's query is "*Politics in Jammu*". In the query there are two terms Politics and Jammu.

**[Table 3]**

| Terms | Info |
|---|---|
| Jammu | {3,<B>,B4},{3,<H>,B5} |
| Politics | {7,<B>,B1} |

Table: 3. Result of query "politics in Jammu

**[Table 4]**

| F1 | F2 | F3 | F4 | F5 | F6 | F7 | F8 |

| Blog id | #subs | #Valid comments | #votes | #likes | #rating | #sharing | #tags | Blogger info |
|---|---|---|---|---|---|---|---|---|
| B1 | 2 | 0 | 4 | 0 | 0 | 0 | 4 | 1 |
| B4 | 14 | 10 | 20 | 3 | 3 | 2 | 5 | 1 |
| B5 | 12 | 3 | 0 | 0 | 0 | 0 | 4 | 0 |

**Table: 4. User Feedback Index**

The blog pages $B_1$, $B_4$, $B_5$ are returned along with the associated information. Let the weight assigned to the query factor is 0.6. For ranking these blog pages,

Calculating score for B1, B1 contains only 1 terms ie "Politics" in the body section (weight assigned=1) and the term appear 7 times.

So score for B1=7+1=8

Taking average B1s=8/2=4

Multiplying by weight ie 0.6 = 4x0.6=2.4.

B1s=2.4……………………………….(i)

Next step is to calculate the score using Table 4. Table 4 contains various feedbacks given by the user on these blogs. Score given on the user feedback is the summation of all feedback factors.

For B1, Considering F1 to F6

Score B1fs= 2+0+4+0+0+0=6

Calculating average, 6/6=1

Multiplying by weight=0.3

Score B1fs= 0.3

Considering F7 and F8

Score B1fs= 4+1=5

Calculating average, 5/2= 2.5

Multiplying by its weight i.e. 0.1= 2.5x0.1=0.25

Considering F1 to F8,

Score B1fs= 0.3+0.25= 0.55………….. (ii)

Considering both the major factors, Query factor and user feedback factor, score of blog page B1

From (i) and (ii)

$S_{B1}$=B1s + B1fs

 2.4 + 0.55=2.95

Where $S_{B1}$ is the score of Blog page B1

Calculating score for Blog B4

For B4, B4 contains only 1 term ie "Jammu" in the body section (weight assigned=1) and the term appears 3 times.

So score for B4=3+1=4

Taking average B4s=4/2=2

Multiplying by weight ie 0.6 = 2x0.6=1.2

B4s=1.2………………………………. (i)

Next step is to calculate the score using Table 4. Score given on the user feedback is the summation of all feedback factors.

For B4, Considering F1 to F6

Score B4fs= 14+10+20+3+3+2=52

Calculating average, 52/6=8.6

Multiplying by weight i.e. 0.3= 8.6x0.3=2.58

Score B4fs= 2.58

Considering F7 and F8

Score B4fs= 5+1=6

Calculating average, 6/2= 3

Multiplying by its weight i.e. 0.1= 3x 0.1=0.3

Considering F1 to F8,

Score B4fs= 2.58+0.3=2.88………….. (ii)

Considering both the major factors, Query factor and user feedback factor, score of blog page B4

From (i) and (ii)

$S_{B4}$=B4s + B4fs

   1.2 + 2.88=4.08

Where $S_{B4}$ is the score of Blog page B4

Calculating score for Blog B5

For B5, B5 contains only 1 terms i.e. "Jammu" in the Heading section (weight assigned=2) and the term appears 3 times.

So score for B5=3+2=5

Taking average B5s=5/2=2.5

Multiplying by weight i.e. 0.6 = 2.5x0.6=1.5

B5s=1.5………………………………. (i)

Next step is to calculate the score using Table 4. Table 4 contains various feedbacks given by the user on the blogs. Score given on the user feedback is the summation of all feedback factors.

For B5, Considering F1 to F6

Score B5fs= 12+3+0+0+0+0=15

Calculating average, 15/6=2.5

Multiplying by weight i.e. 0.3= 2.5x0.3=0.75

Score B5fs= 0.75

Considering F7 and F8

Score B5fs= 4+0=4

Calculating average, 4/2= 2

Multiplying by its weight i.e. 0.1= 2x 0.1=0.2

Considering F1 to F8,

Score B5fs= 0.75+0.2= 0.95………….. (ii)

Considering both the major factors, Query factor and user feedback factor, score of blog page B1,From (i) and (ii)

$S_{B5}$=B5s + B5fs= 1.5 + 0.95=2.45

Where $S_{B5}$ is the score of Blog page B5

A score has been assigned to each blog page and then the blog pages are sorted in decreasing order of their scores. The sorted list of Blog pages is then presented to the user in response to his query with most relevant at the top. Accordingly, Blog 4 is ranked highest with a score of 4.08, followed by Blog 1 at second position with a score of 2.95 and lastly Blog 5 with a score of 2.45.

# [IV]EXPERIMENTAL EVALUATION

The system has been implemented in java using SQL Server 5.0. The snapshots of the implementation have been shown in Fig. 1 and Fig. 2 for the user's query "musical instruments" and "types of musical instruments" respectively. Extensive experiments were conducted over several sources on the blogosphere with the goal to evaluate accuracy of the proposed architecture. The system has shown fairly general behavior with consistent results.

Fig 2 and Fig. 3 shows the number of results retrieved by the system when different queries are entered.

Fig :2. List of blog pages retrieved for the query "musical instrument"

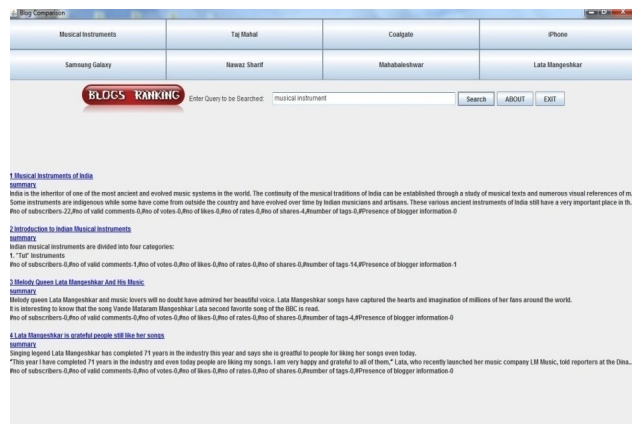

Fig :3. List of blog pages retrieved for the query "types of musical instruments"

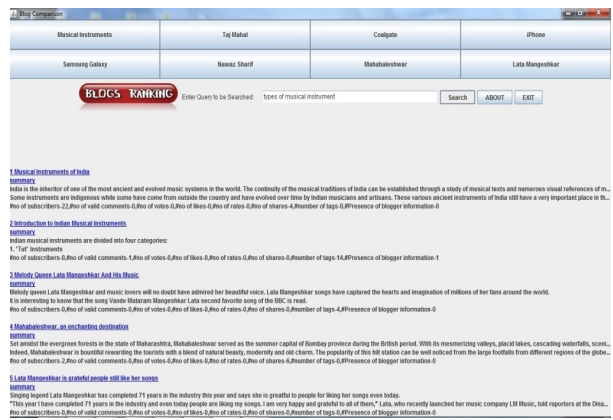

The proposed architecture collected about 180 blog pages from different blog sits covering variety of topics. After collecting the blog pages, analysis was done by applying performance metrics on the sample. To perform the experiments, various performance metrics were taken into consideration. The results obtained thereof from the proposed system were found to be promising. The performance metrics are discussed as follows:

### 4.1. Performance Metrics

To evaluate the proposed work, the three performance metrics namely *precision, recall and F-measure* [10] have been used. Before discussing the performance metrics in detail, let us at first, define the following:

- number of relevant blog pages returned by the system are RB,
- the number of irrelevant pages returned by the system are IB and
- the number of blog pages relevant blog pages that are not returned by system are NB.

With the help of above defined terms RB, IB and NB, let us now further define the various performance metrics:

1. *Precision* is defined as a fraction of relevant blog pages returned by the system. Mathematically, the *Precision* is given by

$P = RB/ (RB + IB)$

2. *Recall* is defined as a fraction of relevant blog pages returned by the proposed system over all the relevant blog pages as given by

experts, then the *Recall* of the proposed system is given by the expression:

$R = RB/(RB + NB)$

3. *F-measure* incorporates both precision and recall. *F-measure* is given by

$F = 2PR/(P + R)$

Where Precision P and Recall R are equally weighted. The values of precision, recall of the system is shown with the help of a graph in Fig.4. It can be seen that *precision o*f the proposed system ranges from 80.51% to 92.1%, *Recall* ranges from 81.66% to 93.2% and *F-measure* ranges from 81.12% to 93.03%. Comparing these values with those of existing systems, the values are found to be higher. Fig. 4 shows precision and recall of the system

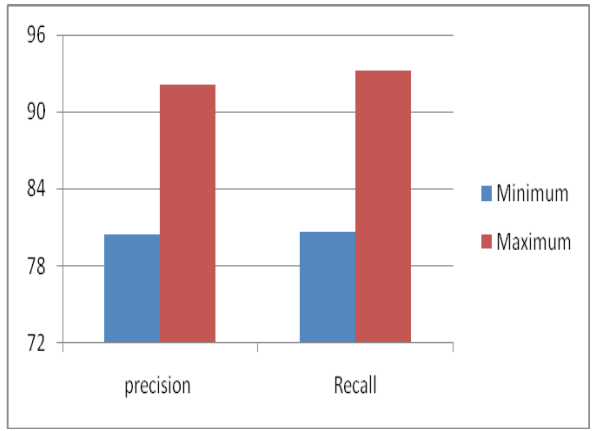

**Fig 4: Precision and recall graph**

The results obtained thereof from the proposed system were found to be promising.

 **[V] CONCLUSION**

The paper presents a new approach in ranking the blog pages. The blog pages are summarized for the purpose of filtering out the non relevant content from the blog pages The relevant content representing the whole page is kept as summary of the blog, which is then indexed and is used for responding to user's query. Along with this, some factors are identified for identifying rank for each page. The higher the rank, the more relevant the web page is. For the user's query, the list of sorted blog pages is then given to the user such that, blog pages with the highest rank appear at the top.

### REFERENCES

[1] BIHUN, Andriy, et al. "Ranking blog documents." WIPO Patent No. 2007033202. 23 Mar. 2007.

[2] Dinesh Sharma, A.K. Sharma, Komal Kumar Bhatia*, "Search engines: a comparative review"*, Proc. of NGCIS-2007.

[3] Das, Dipanjan, and André FT Martins. "*A survey on automatic text summarization.*" *Literature Survey for the Language and Statistics II course at CMU* 4 (2007): 192-195.

[4] Hovy, Eduard, and Daniel Marcu. "*Automated text summarization.*" *The Oxford Handbook of computational linguistics* (2005): 583-598.

[5] Justus Bross,Keven Richly,Matthias Cohnen,Christoph Meinel, "*Identifying the Top dogs of Blogosphere"*, Springer-Verlag 2011

**[6]** Kritikopoulos, Apostolos, Martha Sideri, and Iraklis Varlamis. "*BLOGRANK: Ranking on the blogosphere.*" *Proceedings of the International Conference on Weblogs and Social Media (ICWSM 2007), Boulder, Colorado, USA*. 2007.

**[7]** Madaan, Rosy, A. K. Sharma, and Ashutosh Dixit. "*Presence Factor-Oriented Blog Summarization*.", DBLP (2013).

**[8]** Madaan, Rosy, Ashok Sharma, and Ashutosh Dixit. "*A novel architecture for a blog crawler." Parallel Distributed and Grid Computing (PDGC), 2012 2nd IEEE International Conference on*. IEEE, 2012.

**[9]** Shen, Jie, et al. "*A content-based algorithm for blog ranking.*" *Internet Computing in Science and Engineering, 2008. ICICSE'08. International Conference on*. IEEE, 2008.

**[10]** Sergei Brin and Lawrence Page, "*The anatomy of a large-scale hypertextual Web search engine", Computer Networks and ISDN Systems, 30(1–7):107–117, April 1998*

**[11]** Tayebi, Mohammad A., S. Mehdi Hashemi, and Ali Mohades. "*B2Rank: An algorithm for ranking blogs based on behavioral features.*" *Web Intelligence, IEEE/WIC/ACM International Conference on*. IEEE, 2007

**[12]** Terrence A. Brooks, "*Web Search: How the Web has changed information retrieval", Information Research, April 2003.*